\documentclass[oribibl]{llncs}
\usepackage{multicol}
\usepackage{amsmath}
\usepackage{amssymb}
\usepackage{graphicx}
\usepackage{color}
\usepackage{hhline}
\usepackage{times}

\newcommand{\ignore}[1]{}

\newcommand{\mf}[1]{\mathfrak{ #1}}

\newcommand{\e}{\mathbf{e}}
\newcommand{\msf}[1]{\mbox{\sf #1}}
\newcommand{\msfs}[1]{\mbox{\sf \small #1}}

\newcommand{\red}[1]{\textcolor{red}{#1}}

\newcommand{\bblue}[1]{#1}
\newcommand{\re}[1]{#1}
\newcommand{\bl}[1]{#1}

\newcommand{\ul}[1]{\underline{ #1}}

\newcommand{\boxtheorem}{  \hfill $\Box$}
\newcommand{\nit}[1]{{\it #1}}

\newcommand{\mc}[1]{\mathcal{ #1}}

\newcommand{\Resp}{\nit{Resp}}
\newcommand{\Shap}{\nit{Shap}}
\newcommand{\mbb}[1]{\mathbb{ #1}}

\newcommand{\bcq}{BCQ}
\newcommand{\cq}{CQ}

\newcommand{\fo}{FO}

\title{Attribution-Scores in Data Management and Explainable Machine Learning\thanks{Paper associated to tutorial at ADBIS 2023.}}

\author{{\bf Leopoldo Bertossi\vspace{-2mm}}\thanks{Email: leopoldo.bertossi@skema.edu. \ Member of the Millennium Inst. for Foundational \linebreak Research on Data (IMFD, Chile)}}
\institute{\bf SKEMA Business School, Montreal, Canada \vspace{-3mm}}

\begin{document}

\thispagestyle{empty}
\pagestyle{plain}
\maketitle

\begin{abstract}
We describe recent research on the use of actual causality in the definition of responsibility scores as explanations for query answers in databases, and for outcomes from classification models in machine learning. In the case of databases, useful connections with database repairs are illustrated and exploited. Repairs are also used to give a quantitative measure of the consistency of a database.  For classification models, the responsibility score is properly extended and illustrated. The efficient computation of Shap-score is also analyzed and discussed. The emphasis  is placed on work done by the author and collaborators.
\end{abstract}

\section{Introduction}

\vspace{-2mm}
In data management and artificial intelligence, and machine learning in particular, one wants {\em explanations} for certain results. For example, for query answers and inconsistency in databases. In machine learning (ML), one wants explanations for automated classification results, and automated decision making. Explanations, that may come in different forms, have been the subject of philosophical enquires for a long time, but, closer to our discipline, they appear under different forms in model-based diagnosis, and in causality as developed in artificial intelligence.

In the last few years, explanations that are based on {\em numerical scores} assigned to elements of a model that may contribute to an outcome have become popular. These {\em attribution scores} attempt to capture the degree of contribution of those components to an outcome, e.g. answering questions like these: What is the contribution of this tuple to the answer to this query? Or, what is the contribution of this feature value for the label assigned to this input entity?

In this article, we survey some of the recent advances on the definition,  use and computation of the above mentioned  score-based explanations for query answering in databases, and for outcomes from classification models in ML. This is not intended to be an exhaustive survey of the area. Instead, it is heavily influenced by our latest research. \ignore{ Taking advantage of  repairs, we also show how to specify and compute a numerical measure of inconsistency of database \cite{lpnmr19}. In this case, this would be  a {\em global} score, in contrast with the {\em local} scores applied to individual tuples in a database or feature values in an entity. \ }
 To introduce the concepts and techniques we will use mostly examples, trying  to convey the main intuitions and issues.

Different scores have been proposed in the literature, and some that have a relatively older history have been applied. Among the latter we find the general {\em responsibility score} as found in {\em actual causality} \cite{HP05,CH04,halpern15}. For a particular kind of application, one has to define the right  causality setting, and then apply the responsibility measure to the participating variables.

In data management,  responsibility has been used to
quantify the strength of a tuple  as a cause for a query result \cite{suciu,tocs}. \ The responsibility score, $\nit{Resp}$,  is
based on the notions of {\em counterfactual intervention} as appearing in actual causality. More specifically,
(potential) executions of   {\em counterfactual interventions} on a {\em structural logico-probabilistic model} \ \cite{HP05} are considered, with the purpose of answering hypothetical  questions of the form: \ {\em What would happen if we change ...?}.

Database repairs are commonly used to define and obtain semantically correct query answers from a database (DB) that may fail to satisfy a given set of integrity constraints (ICs) \cite{bertossiSynth}. A connection between repairs and actual causality in DBs has been used to obtain complexity results and algorithms for  responsibility  \cite{tocs} (see Section \ref{sec:dbs}).  Actual causality and responsibility can also be defined at the attribute-level in DBs (rather than at the tuple-level)  \cite{kais}. We briefly describe this alternative in Section \ref{sec:causAttr}. \ On the basis  of database repairs, a measure (or global score) to quantify the degree of inconsistency of a DB was introduced in \cite{lpnmr19}. We give the main ideas in Section \ref{sec:inco}.

The \Resp \ score has also been applied to define scores for binary attribute values in classification \cite{tplp}. However, it has to be generalized when dealing with non-binary features \cite{deem}. We describe this generalization in Section \ref{sec:genResp}.

The Shapley value  of {\em coalition game theory} \cite{S53,R88} can be (and has been) used to define attribution scores, in particular in DBs. The main idea is that {\em several tuples together}, much like
{players in a coalition game}, are necessary to  produce a query result. Some may contribute more than others to the {\em  wealth distribution function} (or simply,  game function), which in this case becomes the query result, namely $1$ or $0$ if the query is Boolean, or a number in the case of an aggregation query. This use of Shapley value was developed in \cite{LBKS20,SigRec21}. See also \cite{SigRec23} for a more recent and general discussion of the use of the Shapley value in DBs.

  The Shapley value has also been used to define explanation scores to feature values in ML-based classification, taking the form of the \Shap \ score \cite{LL17}. Since its computation  is bound to be intractable in general, there has been recent research on classes of models for which \Shap \ becomes tractable \cite{AAAI21,AAAI21ext,guyAAAI21ext}. See Section \ref{sec:shapTract}.

  There hasn't been much research on the use or consideration of domain knowledge (or domain semantics) in the definition and computation of attribution scores. In Section \ref{sec:last}, we describe some problems that emerge in this context.

  This article has \cite{bda22} as a companion, which concentrates mostly on data management. It delves into  the  \Resp \ score under ICs, and on the use of the
  Shapley value for query answering in DBs.  That paper also discusses the
  {\em Causal Effect} score applied in data management \cite{tapp16}. It is also based on causality, as it appears mainly in  {\em observational studies} \cite{rubin,holland,pearl}.

\vspace{-2mm}
\section{Causal Responsibility in Databases}\label{sec:dbs}

\vspace{-2mm}
 Before going into the subject, we recall some notions and notation used in data management. \  A relational schema $\mc{R}$ contains a domain of constants, $\mc{C}$,  and a set of  predicates of finite arities, $\mc{P}$. \ $\mc{R}$ gives rise to a language $\mf{L}(\mc{R})$ of first-order (FO)  predicate logic with built-in equality, $=$.  Variables are usually denoted with $x, y, z, ...$; and finite sequences thereof with $\bar{x}, ...$; and constants with $a, b, c, ...$, etc. An {\em atom} is of the form $P(t_1, \ldots, t_n)$, with $n$-ary $P \in \mc{P}$   and $t_1, \ldots, t_n$ {\em terms}, i.e. constants,  or variables.
  An atom is {\em ground} (a.k.a. a tuple) if it contains no variables. A database (instance), $D$, for $\mc{R}$ is a finite set of ground atoms; and it serves as an  \ignore{The {\em active domain} of a database $D$, denoted ${\it Adom}(D)$, is the set of constants that appear in atoms of $D$.} interpretation structure for  $\mf{L}(\mc{R})$.

A {\em conjunctive query} (\cq) is a \fo \ formula,  $\mc{Q}(\bar{x})$, of the form \ $\exists  \bar{y}\;(P_1(\bar{x}_1)\wedge \dots \wedge P_m(\bar{x}_m))$,
 with $P_i \in \mc{P}$, and (distinct) free variables $\bar{x} := (\bigcup \bar{x}_i) \smallsetminus \bar{y}$. If $\mc{Q}$ has $n$ (free) variables,  $\bar{c} \in \mc{C}^n$ \ is an {\em answer} to $\mc{Q}$ from $D$ if $D \models \mc{Q}[\bar{c}]$, i.e.  $Q[\bar{c}]$ is true in $D$  when the variables in $\bar{x}$ are componentwise replaced by the values in $\bar{c}$. $\mc{Q}(D)$ denotes the set of answers to $\mc{Q}$ from $D$. $\mc{Q}$ is a {\em Boolean conjunctive query} (\bcq) when $\bar{x}$ is empty; and when {\em true} in $D$,  $\mc{Q}(D) := \{\nit{true}\}$. Otherwise, it is {\em false}, and $\mc{Q}(D) := \emptyset$. We will consider only conjunctive queries, which are the most common the data management.

Integrity constraints (ICs) are sentences of $\mf{L}(\mc{R})$ that a DB is expected to satisfy. Here, we consider mainly  {\em denial constraints} \ (DCs), i.e.  of the form $\kappa\!:  \neg \exists \bar{x}(P_1(\bar{x}_1)\wedge \dots \wedge P_m(\bar{x}_m))$,
where $P_i \in \mc{P}$, and $\bar{x} = \bigcup \bar{x}_i$.
\  If an instance $D$ does not satisfy the set $\Sigma$ of ICs associated to the schema, we say that $D$ is {\em inconsistent}, denoted with \ $D \not \models \Sigma$.

Now we move into the proper subject of this section. \  In data management we {need to understand and compute}
{\em  why}  certain results are obtained or not, e.g. query answers,  violations of semantic conditions, etc.; and we
expect a  database system to provide {\em explanations}. \ Here, we will consider explanations that are based on {\em actual causality} \cite{HP05,halpern15}. They were introduced in
 \cite{suciu,suciuDEBull}, and will be illustrated by means of an example.

 \begin{example}  \label{ex:uno}  \ Consider the database ${D}$, and the Boolean conjunctive query (BCQ)

\begin{multicols}{2}

\vspace{4mm}\hspace*{1cm}\begin{tabular}{l|c|c|} \hline
$R$  & $A$ & $B$ \\\hline
 & $a$ & ${b}$\\
& $c$ & $d$\\
& ${b}$ & ${b}$\\
 \hhline{~--}
\end{tabular} \hspace*{0.5cm}\begin{tabular}{l|c|c|}\hline
$S$  & $C$  \\\hline
 & $a$ \\
& $c$ \\
& ${b}$ \\ \hhline{~-}
\end{tabular}

\phantom{oo}

\phantom{oo}

\vspace{-6mm}
\noindent $\mc{Q}\!: \ \exists x \exists y ( S(x) \land R(x, y) \land S(y))$,
for which \ ${D \models \mc{Q}}$ holds, i.e. the query is true in $D$. \ We ask about the  causes for $\mc{Q}$ to be true. \ A tuple ${\tau \in D}$ is
{\em counterfactual cause} for  ${\mc{Q}}$ \ (being true in $D$) \ if \ ${D\models \mc{Q}}$ 

\phantom{oo}
\end{multicols}

\vspace{-7mm}  \noindent and \ ${D\smallsetminus \{\tau\}  \not \models \mc{Q}}$.
\ In this example,   {$S(b)$ is a counterfactual cause for $\mc{Q}$}: \ If ${S(b)}$ is removed from ${D}$,
 ${\mc{Q}}$ is no longer true.

Removing a single tuple may not be enough to invalidate the query. Accordingly, a tuple ${\tau \in D}$ is  an {\em actual cause} for  ${\mc{Q}}$
if there  is a {\em contingency set} \ ${\Gamma \subseteq (D \smallsetminus \{\tau\})}$,  such that \ ${\tau}$ \ is a   counterfactual cause for ${\mc{Q}}$ in ${D\smallsetminus \Gamma}$. That is, $D \models \mc{Q}$, \ $D \smallsetminus \Gamma \models \mc{Q}$, but $D \smallsetminus (\Gamma \cup \{\tau\}) \not \models \mc{Q}$.
\ In this example,  ${R(a,b)}$ is not a counterfactual cause for ${\mc{Q}}$, but it is an actual cause  with contingency set
${\{ R(b,b)\}}$: \ If ${R(b,b)}$ is removed from ${D}$, ${\mc{Q}}$ is still true, but further removing ${R(a,b)}$ makes ${\mc{Q}}$ false.
\boxtheorem \end{example}

Notice that every counterfactual cause is also an actual cause, with empty contingent set.   Actual causes that are not counterfactual causes need company to invalidate a query result.
 \ Now, we ask  how strong are tuples as actual causes. \ To answer  this question, we appeal to the {\em responsibility} of an actual cause ${\tau}$ for ${\mc{Q}}$ \cite{suciu}, defined by:

 \vspace{-2mm}
\begin{equation*}
{\nit{Resp}_{\!_D}^{\!\mc{Q}}\!(\tau) \ := \ \frac{1}{|\Gamma| \ + \ 1}},
\end{equation*}
where ${|\Gamma|}$ is the
size of a smallest contingency set, $\Gamma$, for ${\tau}$, \ and  $0$, otherwise.

\vspace{-2mm}
\begin{example} \ (ex. \ref{ex:uno} cont.) \ The {responsibility of ${R(a,b)}$ is \  $\frac{1}{2}$} ${= \frac{1}{1 + 1}}$ \ (its several smallest contingency sets have all size ${1}$). \
  ${R(b,b)}$ and ${S(a)}$ are also actual causes with responsibility  \ ${\frac{1}{2}}$; and
  ${S(b)}$ is actual (counterfactual) cause with responsibility \   $1$ ${= \frac{1}{1 + 0}}$. \boxtheorem
\end{example}

We can see that causes are, in this database context, tuples that come with their responsibilities as  ``scores". \ It turns out that there are precise connections between {\em database repairs} and tuples as actual causes for queries answers in databases. These connections where exploited to obtain complexity results for responsibility \cite{tocs} (among other uses, e.g. to obtain answer-set programs for the specification and computation of actual causes and responsibility \cite{kais}).

The notion of {\em repair} of a relational database was introduced in order to formalize the notion of {\em consistent query answering} (CQA)  \cite{pods99,bertossiSynth}: If a database $D$ is inconsistent in the sense that is does not satisfy a given set of integrity constraints, $\nit{ICs}$, and a query $\mc{Q}$ is posed to $D$, what are the meaningful, or consistent, answers to $\mc{Q}$ from $D$? They are sanctioned as those that hold (are returned as answers) from {\em all} the {\em repairs} of $D$. The repairs of $D$ are consistent instances $D'$ (over the same schema of $D$), i.e. $D' \models \nit{ICs}$, and {\em minimally depart} from $D$.

\begin{example} \label{ex:theEx} Let us consider the following set of {\em denial constraints} (DCs) and a database $D$, whose relations (tables) are shown right here below. $D$ is inconsistent, because it violates the DCs:  it satisfies the joins that are prohibited by the DCs.

\vspace{-6mm}
\begin{multicols}{2}
\begin{eqnarray*}
\neg \exists x \exists y(P(x) \wedge Q(x,y))\\
\neg \exists x \exists y(P(x) \wedge R(x,y))
\end{eqnarray*}

\phantom{ooo}

\begin{tabular}{c|c|}\hline
$P$&A\\ \hline
&a\\
&e\\ \hhline{~-}
\end{tabular}~~~~~~~~
\begin{tabular}{c|c|c|}\hline
$Q$&A&B\\ \hline
& a & b\\ \hhline{~--}
\end{tabular}~~~~~~~~
\begin{tabular}{c|c|c|}\hline
$R$&A&C\\ \hline
& a & c\\ \hhline{~--}
\end{tabular}
\end{multicols}

\vspace{-2mm}We want to repair the original instance by {\em deleting tuples} from relations. Notice that, for DCs, insertions of new tuple will not restore consistency. We could change (update) attribute values though, a possibility that has been investigated in \cite{kais}.\ignore{will consider in Section \ref{sec:causAttr}.}

Here we have two {\em subset-repairs}, a.k.a. {\em S-repairs}. They are subset-maximal consistent subinstances of $D$: \ $D_1 = \{P(e), Q(a,b), R(a,c)\}$ \ and \ $D_2 = \{P(e),$ $ P(a)\}$. They are consistent, subinstances of $D$, and any proper superset of them (still contained in $D$) is inconsistent. (In general, we will represent database relations as set of tuples.)

We also have {\em cardinality repairs}, a.k.a.  {\em C-repairs}. They are consistent subinstances of $D$ that minimize the {\em number} of tuples by which they differ from $D$. That is, they are maximum-cardinality consistent  subinstances. In this case, only
$D_1$ is a C-repair. \ Every C-repair is an S-repair, but not necessarily the other way around. \boxtheorem
\end{example}

Let us now consider a BCQ
\begin{equation}
\mc{Q}\!: \exists \bar{x}(P_1(\bar{x}_1) \wedge \cdots \wedge P_m(\bar{x}_m)),
\end{equation}
which we assume is  true in  a database $D$. \ It turns out that we can obtain the causes for $\mc{Q}$ \ to be true $D$, and their contingency sets from database repairs. In order to do this, notice that
$\neg \mc{Q}$ \ becomes a  DC \begin{equation}
\kappa(\mc{Q})\!: \   \neg \exists \bar{x}(P_1(\bar{x}_1) \wedge \cdots \wedge P_m(\bar{x}_m)).
\end{equation}

$\mc{Q}$ holds in $D$ \  iff \  $D$ is inconsistent w.r.t. $\kappa(\mc{Q})$.

S-repairs are associated to causes with minimal contingency sets, while C-repairs are associated to causes for $\mc{Q}$ with minimum contingency sets, and maximum responsibilities \cite{tocs}. In fact,
for a database tuple \ $\tau \in D$:
\begin{itemize}
\item [(a)] $\tau$ \ is actual cause for $\mc{Q}$ with subset-minimal contingency set $\Gamma$ \ iff \ $D \smallsetminus (\Gamma \cup \{\tau\})$ \ is an  S-repair (w.r.t. $\kappa(\mc{Q})$),
in which case, its responsibility is \ $\frac{1}{1 + |\Gamma|}$. \item[(b)]
$\tau$ \ is actual cause with minimum-cardinality contingency set $\Gamma$ \ iff \ $D \smallsetminus (\Gamma \cup \{\tau\})$ \ is C-repair,
in which case, $\tau$ is a maximum-responsibility actual cause.
\end{itemize}
Conversely, repairs can be obtained from causes and their contingency sets \cite{tocs}. These results can be extended to unions of BCQs (UBCQs), or equivalently, to sets of denial constraints.

One can exploit the connection between causes and repairs to understand the computational complexity of the former by leveraging existing results for the latter. Beyond
the fact that computing or deciding actual causes  can be done in polynomial time  in data for CQs and unions of CQs (UCQs) \cite{suciu,tocs}, one can show that
most computational problems related to responsibility are intractable, because they are also intractable for repairs, in particular,  for C-repairs (all this in data complexity)
\cite{lopatenko}. In particular, one can prove \cite{tocs}:
\begin{itemize}
\item[(a)] The {\em responsibility problem}, about deciding if a tuple has responsibility above a certain threshold, is $\nit{N\!P}$-complete for  UCQs. \ However, on the positive side, this problem is {\em fixed-parameter tractable} (equivalently, belongs to the {\em FPT} class) \cite{flum}, with the parameter being the inverse of the threshold. \item[(b)] Computing  $\nit{Resp}_{_{\!D\!}}^{\!\mc{Q}}(\tau)$ \ is $\nit{F\!P}^{\nit{N\!P(log} (n))}$-complete for BCQs. This the {\em functional}, non-decision, version of the  responsibility problem. The complexity class involved is that of computational problems that use polynomial time with a logarithmic number of calls to an oracle in \nit{NP}. \item[(c)]
Deciding if a tuple $\tau$ is a most responsible cause is  $P^\nit{N\!P(log(n))}$-complete for BCQs. The complexity class is as the previous one, but for decision problems \cite{arora}.
\end{itemize}

For further illustration, property (b) right above tells us that there is a database schema and a Boolean conjunctive query $\mc{Q}$, such that computing the responsibility of tuple in an instance $D$ as an actual cause for $\mc{Q}$ being true in $D$ is $\nit{F\!P}^{\nit{N\!P(log} (n))}$-hard in $|D|$. This is due to the fact that, through the C-repair connection,
determining the responsibility of a tuple becomes the problem on hyper-graphs of determining the size of a minimum vertex cover that contains the tuple as a vertex (among all vertex covers that contain the vertex) \cite[sec. 6]{tocs}. This latter problem was first investigated in the context of C-repairs w.r.t. denial constraints. Those repairs  can be characterized in terms of hyper-graphs \cite{lopatenko} (see Section \ref{sec:inco} for a simple example). \ This hyper-graph connection also allows to obtain the FPT result in (a) above, because we are dealing with hyper-edges that have a fixed upper bound given by the number of atoms in the denial constraint associated to que conjunctive query $\mc{Q}$ (see \cite[sec. 6.1]{tocs} for more details).

Notice that a class of database repairs determines a {\em possible-world semantics} for database consistency restoration. We could use in principle any reasonable notion of distance between database instances, with each choice defining a particular {\em repair semantics}. S-repairs and C-repairs are examples of repair semantics. By choosing alternative repair semantics and an associated notion of counterfactual intervention, we can {\em define} also alternative notions of actual causality in databases and associated responsibility measures \cite{kais}.
\ We will show an example in Section \ref{sec:causAttr}.

\vspace{-2mm}
\subsection{Attribute-Level Causal Responsibility in Databases}\label{sec:causAttr}

\vspace{-2mm}
Causality and responsibility in databases can be extended to the attribute-value level; and can also be connected with appropriate forms of database repairs \cite{tocs,kais}.
\ We will develop this idea in the light of an example, with a particular repair-semantics, and we will apply it to define attribute-level causes for query answering, i.e. we are interested in attribute values in tuples rather than in whole tuples. \ The repair semantics we use here is very natural, but others could be used instead.

\begin{example} \label{ex:atCaus} \ Consider the database $\bblue{D}$, with tids,  and query \ $\bblue{\mc{Q}\!: \ \exists x \exists y ( S(x) \land R(x, y) \land S(y))}$, of Example \ref{ex:uno} \ and  \ the associated denial constraint \ $\kappa(\mc{Q}): \ \neg \exists x\exists y( S(x)\wedge R(x, y)\wedge S(y))$.

\vspace{-2mm}
\begin{multicols}{2}

\hspace*{1cm}\re{\begin{tabular}{l|c|c|} \hline
$R$  & A & B \\\hline
$t_1$ & $a$ & $b$\\
$t_2$& $c$ & $d$\\
$t_3$& $b$ & $b$\\
 \hhline{~--}
\end{tabular} \hspace*{0.5cm}\begin{tabular}{l|c|c|}\hline
$S$  & C  \\\hline
$t_4$ & $a$ \\
$t_5$& $c$ \\
$t_6$ & $b$\\
 \hhline{~--}
\end{tabular}  }

\noindent Since $\bblue{D \not \models \kappa(\mc{Q})}$, we need to consider repairs of $D$ w.r.t. $\kappa(\mc{Q})$. \ Repairs will be obtained  by ``minimally" changing attribute  values by {\sf NULL},  as in SQL databases, \linebreak
\end{multicols}

\vspace{-6mm}\noindent which
cannot be used to satisfy a join. In this case, minimality means that {\em the set} of values changed by {\sf NULL} is minimal under set inclusion.  These are two different minimal-repairs:

\vspace{-3mm}
\begin{multicols}{2}
\hspace*{10mm}\begin{tabular}{l|c|c|} \hline
$R$  & A & B \\\hline
$t_1$& $a$ & $b$\\
$t_2$& $c$ & $d$\\
$t_3$& $b$ & $b$\\
 \hhline{~--}
\end{tabular} \hspace*{0.5cm}\begin{tabular}{l|c|c|}\hline
$S$  & C  \\\hline
$t_4$& $a$ \\
$t_5$& $c$ \\
$t_6$& $\re{\sf NULL}$ \\ \hhline{~-}
\end{tabular}

\hspace*{10mm}\begin{tabular}{l|c|c|} \hline
$R$  & A & B \\\hline
$t_1$ & $a$ & $\re{\sf NULL}$\\
$t_2$& $c$ & $d$\\
$t_3$& $b$ & $\re{\sf NULL}$\\
 \hhline{~--}
\end{tabular} \hspace*{0.5cm}\begin{tabular}{l|c|c|}\hline
$S$  & C  \\\hline
 $t_4$& $a$ \\
$t_5$& $c$ \\
$t_6$& $b$ \\ \hhline{~-}
\end{tabular}
\end{multicols}

\vspace{-2mm}It is easy to check that they do not satisfy $\kappa(\mc{Q})$. \  If we denote the changed values by the tid with the position where the changed occurred, the first repair is characterized by the set $\{t_6[1]\}$, whereas the second, by the set $\{t_1[2], t_3[2]\}$. Both are minimal since none of them is contained in the other.

Now, we could also introduce a notion of {\em cardinality-repair}, keeping those where the number of changes is a minimum. In this case, the first repair qualifies, but not the second.
\ These repairs identify (actually, define) the value in $\re{t_6[1]}$ as a maximum-responsibility cause for $\mc{Q}$ being true (with responsibility $1$). Similarly,  \ $\re{t_1[2]}$ and $\re{t_3[2]}$ \ become actual causes, that do need contingent companion values,  which makes them take a responsibility of $\frac{1}{2}$ each. \boxtheorem
\end{example}

 We should emphasize that, under the semantics illustrated with the example, we are considering attribute values participating in joins as interesting causes. A detailed treatment of the underlying repairs semantics with its application to attribute-level causality can be found in \cite{kais}. In a related vein,
one could also consider as causes other attribute values in a tuple that participate in a query (being true), e.g. that in $t_3[1]$, but making them {\em non-prioritized} causes. One could also think of adjusting the responsibility measure in order to give to these causes a lower score.

More generally, in order to define actual causality and responsibility at the attribute level, we could -alternatively- consider all the attributes in tuples (in joins or not), and all the possible  updates drawn from an attribute domain for a given value in a given tuple in the database, so as for potential contingency sets for it. If the attribute domains contain $\re{\sf NULL}$, this update semantics would generalize the one illustrated with Example \ref{ex:atCaus}. As earlier in this section, the purpose of all this would be to counterfactually (and minimally) invalidate the query answer. On the repair counterpart, we could consider {\em update-repairs} \cite{wijsen} to restore the satisfaction of the associated denial constraint.

In this more general situation, where we may depart from the ``binary" case so far (i.e. the tuple is or not in the DB, or a value is replaced by a null or not), we have to be more careful: Some value updates on a tuple may invalidate the query and some other may not. So, what if only one such update invalidates the query, but all the others do not? This would not be reflected in the responsibility score as defined so far. We could generalize the responsibility score by bringing into the picture the average (or expected) value of the query (of $1$s or $0$s for true or false) in the so-intervened database. This idea has been developed in the context of explanation scores for ML-based classification \cite{deem}. See Section \ref{sec:genResp} for details.

\vspace{-2mm}
\section{Measuring Database Inconsistency}\label{sec:inco}
\vspace{-2mm}

A database $D$ is expected to satisfy a given set of integrity constraints (ICs), $\Sigma$, that come with the database schema. However, databases may be inconsistent in that those ICs are not satisfied. A natural question is: \
{\em To what extent, or how much inconsistent is \ $D$ \ w.r.t. \ $\Sigma$, in quantitative terms?}. \ This problem is about defining a {\em global numerical score} for the database, to capture its ``degree of inconsistency". This number can be interesting {\em per se}, as a measure of data quality (or a certain aspect of it), and could also be used to compare two databases (for the same schema) w.r.t. (in)consistency.

In \cite{lpnmr19}, a particular and natural  {\em inconsistency measure} (IM) was introduced and investigated. More specifically, the measure was inspired by the $g_3$-measure used for functional dependencies (FDs) in \cite{mannila}, and reformulated and generalized {\em in terms of a class of database repairs}. \ Once a particular measure is defined,  an interesting research program can be developed on its basis. In particular, around its mathematical properties,  and the algorithmic and complexity issues related to its computation. \
In \cite{lpnmr19}, in  addition to algorithms, complexity results, approximations for  hard cases of the IM computation, and the dynamics of the IM under updates,  {\em answer-set programs} were proposed for the computation of  this measure.

In the rest of this section, we will use the notions and notation introduced in  Example \ref{ex:theEx}. \ For a database $D$ and a set of {\em denial constraints} $\Sigma$ (this is not essential, but to fix ideas), we have the classes of subset-repairs (or S-repairs), and cardinality-repairs (or C-repairs), denoted $\nit{Srep}(D,\Sigma)$ and $\nit{Crep}(D,\Sigma)$, resp. \ The consider the following inconsistency measure: \vspace{-2mm}
\begin{equation*}
\re{\mbox{\nit{inc-deg}}^{C\!}(D,\Sigma)} := \frac{|D| - \nit{max}\{ |D'| ~:~D' \in  \re{\nit{Crep}(D,\Sigma)} \}}{|D|}.\label{eq:c}
\end{equation*}

\vspace{-2mm}It is easy to verify that replacing C-repairs by S-repairs gives the same measure.   \ It is clear that
\ $\bl{0 \leq \mbox{\nit{inc-deg}}^{C\!}(D,\Sigma) \leq 1}$, \ with value $\bl{0}$ when $\bl{D}$ consistent. \ Notice that one could use other repair semantics instead of C-repairs \cite{lpnmr19}.

\begin{example} \ (example \ref{ex:theEx} cont.) \ Here, $\bl{\nit{Srep}(D,\Sigma) = \{D_1, D_2 \}}$ \ and \
$\bl{\nit{Crep}(D,\Sigma) = \{D_1 \}}$. It holds:
$\mbox{\nit{inc-deg}}^{C\!}(D,\Sigma) =  \frac{4 - |D_1|}{4} = \frac{1}{4}.$ \boxtheorem
\end{example}

It turns out that complexity and efficient computation results, when they exist, can be obtained via C-repairs, for which we end up
confronting graph-theoretic problems. Actually, C-repairs are in one-to-one correspondence with maximum-size independent sets in hyper-graphs, whose vertices are the database tuples, and hyper-edges are formed by tuples that jointly violate an IC   \cite{lopatenko}.

\begin{example} \ Consider the database $\bl{D = \{A(a), B(a), C(a), D(a), E(a)\}}$, which is inconsistent w.r.t. the set of DS:

\vspace{-6mm}
$$\bl{\Sigma= \{\neg \exists x(B(x)\wedge E(x)), \ \neg \exists x(B(x) \wedge C(x) \wedge D(x)), \ \neg \exists x(A(x) \wedge C(x))\}}.$$

\vspace{-2mm}We obtain the following {\em conflict hyper-graph} \ (CHG), where tuples become the nodes, and a hyper-edge connects tuples that together violate a DC:

\vspace{-3mm}
\begin{multicols}{2}
\includegraphics[width=3.2cm]{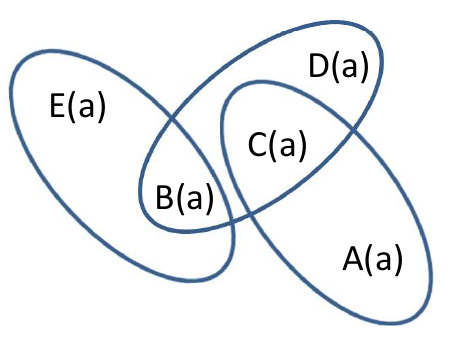}

S-repairs  are maximal \re{independent sets}: \  $\bl{D_1 = \{B(a), C(a)\}}$, \ $\bl{D_2 = \{C(a), D(a),E(a)\}}$, \ \ $\bl{D_3 = \{A(a),B(a), D(a)\}}$; and
the C-repairs are \ $\bl{D_2, \ D_3}$. \boxtheorem
\end{multicols}
\end{example}

\vspace{-5mm}There is a connection between C-repairs and {\em hitting-sets} (HS) of the hyper-edges
of the CHG \cite{lopatenko}: The removal from $\bl{D}$ of the vertices in a minimum-size {\em hitting set} produces a C-repair. \
The connections between hitting-sets in hyper-graphs and C-repairs can be exploited for algorithmic purposes, and to obtain complexity and approximation results \cite{lpnmr19}. \ In particular, the \re{complexity} of computing   \ $\bl{\mbox{\nit{inc-deg}}^{C\!}(D,\Sigma)}$ \ for DCs belongs to  \ $\bl{\nit{FP}^{\nit{NP(log(n))}}}$, in data complexity. Furthermore,
there is a relational schema and a set of DCs $\bl{\Sigma}$ for which
computing \ $\bl{\mbox{\nit{inc-deg}}^{C\!}(D,\Sigma)}$  is \ $\bl{\nit{FP}^{\nit{NP(log(n))}}}$-complete.

Inconsistency measures have been introduced and investigated in knowledge representation, but mainly for  propositional theories \cite{hunter,grant}. In databases, it is more natural to
consider the different nature of  the combination of a database, as  a structure, and ICs, as a set of first-order formulas. It is also important to consider the asymmetry:
databases are inconsistent or not, not the combination. Furthermore, the relevant issues that are usually related to data management have to do with
algorithms  and computational complexity; actually,  in terms of the size of the database (that tends to be by far the largest parameter).

Scores for {\em individual tuples} about their contribution to inconsistency can be obtained through responsibility scores for query answering, because every IC gives rise to a ``violation view" (a query). Also Shapley values have been applied to measure the contribution of tuples to inconsistency of a database w.r.t. FDs \cite{ester}.

\vspace{-2mm}
\section{Attribution Scores in Machine Learning}\label{sec:xai}

\vspace{-2mm}
Scores as explanations for outcomes from machine learning (ML) models have also been proposed. In this section we discuss some aspects of two of them, the \Resp \ and the \Shap \ scores.

\vspace{-1mm}
\begin{example} \label{ex:one} Consider a client of bank who is applying for loan. The bank will process his/her application by means of a ML system that will decide if the client should be given the loan or not. In order for the application to be computationally processed, the client is represented as an {\em entity}, say \ $\e = \langle \msf{\small john}, \msfs{18}, \msfs{plumber}, \msfs{70K},$ $ \msfs{harlem}, \msfs{10K}, \msfs{basic}\rangle$, that is, a finite record
of  {\em feature values}. The set of features is $\mc{F} = \{\msfs{Name}$, $\msfs{Age}$, $\msfs{Activity}$, $\msfs{Income}$, $\msfs{Debt}$, $\msfs{EdLevel}\}$.

The bank uses a {\em classifier}, $\mc{C}$, which,
after receiving input $\e$,  returns a {\em label}: {\em Yes} or {\em No} (or $0$ or $1$). In this case, it returns {\em No}, indicating that the loan request is rejected. \ The client (or the bank executive) asks {\em ``Why?"}, and would like to  have an explanation. \  What kind of explanation? \ How could it be provided? \
From what? \boxtheorem
\end{example}

\vspace{-1mm}There are different ways of building such a classifier, and depending on the resulting model, the classifier may be less or more ``interpretable". For example, complex neural networks are considered to be difficult to interpret, and become ``black-boxes", whereas more explicit models, such as decision trees, are considered to be much more interpretable, and  ``open-boxes" for that matter.

In situations such as that
 in Example \ref{ex:one}, actual causality and responsibility   have been applied to provide {\em counterfactual explanations} for classification results, and scores for them. In order to do this, having access to the internal components of the classifier is not needed, but only its input/output relation.

\begin{example} \label{ex:dos} (example \ref{ex:one} cont.) \ The entity $\e \ = \ \langle \msfs{john}, \msfs{18}, \msfs{plumber}, \msfs{70K}, \msfs{harlem}, \msfs{10K},$ $\msfs{basic}\rangle$
\ignore{\begin{equation}
\e \ = \ \langle \msfs{john}, \msfs{18}, \msfs{plumber}, \msfs{70K}, \msfs{harlem}, \msfs{10K},\msfs{basic}\rangle \label{eq:ent}
\end{equation}}
 received  the label \ $1$ from the classifier, indicating that the loan is not granted.\footnote{We are assuming that classifiers are {\em binary}, i.e. they return labels $0$ or $1$. For simplicity and uniformity, but without loss of generality, we will assume that label $1$ is the one we want to explain.} \ In order to identify counterfactual explanations, we intervene the feature values replacing them by alternative ones from the feature domains, e.g.
\ $\e_1 \ = \  \langle \msfs{john}, \ul{\msfs{25}}, \msfs{plumber},$ $ \msfs{70K},$ $\msfs{harlem},\msfs{10K}, \msfs{basic}\rangle$, which receives the label  \ $0$.
\ The counterfactual version $\e_2 = \ \langle \msfs{john}, \msfs{18},$ $ \msfs{plumber}, \ul{\msfs{80K}}, \ul{\msfs{brooklyn}}, \msfs{10K}, \msfs{basic}\rangle$ also get label \ $0$. Assuming, in the latter case, that none of the single changes alone switch the label, we could say that $\msfs{Age} = \msfs{25}$, so as $\msfs{Income} = \msfs{70K}$ with contingency $\msfs{Location} = \msfs{harlem}$ (and the other way around) in $\e$ are (minimal) counterfactual explanations, by being actual causes for the label.

We could go one step beyond, and define responsibility scores: \ $\nit{Resp}(\e,\msfs{Age}) := 1$, and $\nit{Resp}(\e,\msfs{Income}) := \frac{1}{2}$ (due to the additional, required, contingent change). This choice does reflect the causal strengths of attribute values in $\e$. However, it could be the case that only by changing the value of $\msfs{Age}$ to $\msfs{25}$ we manage to switch the label, whereas for all the other possible values for $\msfs{Age}$ (while nothing else changes), the label is always {\em No}. It seems more reasonable to redefine responsibility by considering an average involving all the possible labels obtained in this way. \boxtheorem
\end{example}

An  application of the responsibility score similar to that in \cite{suciu,tocs} works fine for explanation scores when features are  {\em binary}, i.e. taking two values, e.g. $0$ or $1$ \cite{tplp}. Even in this case,  responsibility computation can be intractable \cite{tplp}. However, as mentioned in the previous example, when features have more than two values, it makes sense to extend the definition of the responsibility score.

\vspace{-2mm}
\subsection{The Generalized $\mathbf{\nit{Resp}}$  Score}\label{sec:genResp}

\vspace{-2mm}
In \cite{deem}, a generalized $\nit{Resp}$ score was introduced and investigated. We describe it next in intuitive terms, and appealing to Example \ref{ex:dos}.

\begin{enumerate}
\item For an entity $\e$ classified with  label $L(\e) =1$, \ and a feature $F^\star$, whose value $F^\star(\e)$ appears in $\e$, we want a numerical responsibility score $\nit{Resp}(\e,F^\star)$, characterizing the causal strength of $F^\star(\e)$ for outcome $L(\e)$. \ In the example, $F^\star = \msfs{Salary}$, $F^\star(\e) = \msfs{70K}$, and $L(\e) = 1$.

\vspace{1mm}
\item  While we keep the original value for $\msfs{Salary}$ fixed, we start by defining a ``local" score for a fixed contingent assignment $\Gamma := \bar{w}$, with $F^\star \notin \Gamma \subseteq \mc{F}$. \ We define $\e^{\Gamma,\bar{w}} := \e[\Gamma:=\bar{w}]$, the entity obtained from $\e$ by changing (or redefining) its values for features in $\Gamma$, according to $\bar{w}$.

  \vspace{1mm}
    In the example, it could be $\Gamma = \{\msfs{Location}\}$, and $\bar{w} := \langle \msfs{brooklin}\rangle$, a contingent (new) value for $\msfs{Location}$. Then, $\e^{\{\msfs{Location}\},\langle \msfs{brooklin}\rangle} = \e[\msfs{Location}:= \msfs{brooklin}]  =  \langle \msfs{john}, \msfs{25}, \msfs{plumber}, \msfs{70K},$ $\ul{\msfs{brooklin}},\msfs{10K},$ $ \msfs{basic}\rangle$.

    \vspace{1mm}
    We make sure (or assume in the following) that $L(\e^{\Gamma,\bar{w}}) = L(\e) = 1$ holds. \ This is because, being these changes only contingent, we do not expect them to switch the label by themselves, but only and until the ``main" counterfactual change on $F^\star$ is made.

\vspace{1mm}
      In the example, we assume $L(\e[\msfs{Location}:= \msfs{brooklin}])  =   1$. \ Another case could be $\e^{\Gamma^{\prime}\!,\bar{w}^\prime}$\!\!, with $\Gamma^\prime = \{\msfs{Activity}, \msfs{Education}\}$, and $\bar{w}^\prime = \langle \msfs{accountant}, \msfs{medium}\rangle$, with $L(\e^{\Gamma^{\prime}\!,\bar{w}^\prime}) = 1$.

\vspace{1mm}
\item Now, for each of those $\e^{\Gamma,\bar{w}}$ as in the previous item, we consider all the different possible values $v$ for $F^\star$, with the values for all the other features fixed as in $\e^{\Gamma,\bar{w}}$.

    For example, starting from  $\e[\msfs{Location}:= \msfs{brooklin}]$, we can consider $\e_1^{\prime} :=$ \linebreak $ \e[\msfs{Location}:= \msfs{brooklin};\ul{\msfs{Salary} :=\msfs{60K}}]$ \ (which is the same as $\e^{\msfs{Location},\langle\msfs{brooklin}\rangle}[$ $\msfs{Salary}$ $ :=\msfs{60K}]$), obtaining, e.g. $L(\e_1^{\prime}) = 1$. \ However, for $\e_2^{\prime} :=$ $ \e[\msfs{Location}:= \msfs{brooklin};$  $\ul{\msfs{Salary :=80}}]$, we now obtain, e.g. $L(\e_2^{\prime}) = 0$, etc.

\vspace{1mm}
For a fixed (potentially) contingent change $(\Gamma, \bar{w})$ on $\e$, we consider the  difference between the original label $1$ and the expected label obtained by further modifying the value of $F^\star$ (in all possible ways). \ This gives us a {\em local} responsibility score, \  local to $(\Gamma, \bar{w})$: \vspace{-4mm}
\begin{eqnarray}
\nit{Resp}(\e,F^\star,\ul{\Gamma,\bar{w}}) &:=& \frac{L(\e) - \mathbb{E}(~L(\e^\prime)~|~F(\e^\prime) = \ F(\e^{\Gamma,\bar{w}}), \ \forall  F \in (\mc{F}\smallsetminus \{F^\star\}~)}{1 + |\Gamma|} \nonumber\\
&=&\frac{1 - \mathbb{E}(~L(\e^{\Gamma,\bar{w}}[F^\star := v])~|~ v \in \nit{Dom}(F^\star)~)}{1 + |\Gamma|}. \label{eq:star}
\end{eqnarray}

\vspace{-1mm}This local score takes into account the size of the contingency set $\Gamma$.

\vspace{1mm}
We are assuming here that there is a probability distribution over the entity population $\mc{E}$. It could be known from the start, or it could be an empirical distribution obtained from a sample.
As discussed in \cite{deem}, the choice (or whichever distribution that is available) is relevant for the computation of the general $\nit{Resp}$ score, which  involves the local ones (coming right here below).

\vspace{1mm}
\item  Now, generalizing the notions introduced in Section \ref{sec:dbs},  we can say that the value $F^\star(\e)$ is
an {\em actual cause} for label $1$ when, for some $(\Gamma, \bar{w})$, (\ref{eq:star}) is positive: at least one change of value for $F^\star$ in $\e$ changes the label (with the company of $(\Gamma, \bar{w})$).

\vspace{1mm}
When $\Gamma = \emptyset$ (and then, $\bar{w}$ is an empty assignment), and (\ref{eq:star}) is positive, it means that at least one change of value for $F^\star$ in $\e$ switches the label by itself. As before, we can say that $F^\star(\e)$ is a {\em counterfactual cause}. However, as desired and expected, it is not necessarily the case anymore that counterfactual causes (as original values in $\e$) have all the same causal strength: $F_i(\e), F_j(\e)$ could be both counterfactual causes, but with different values for (\ref{eq:star}), for example if changes on the first switch the label ``fewer times" than those on the second.

\vspace{1mm}
\item Now, we can define the global score, by considering the ``best" contingencies $(\Gamma,\bar{w})$, which involves requesting from $\Gamma$ to be of minimum size:
     \vspace{-2mm}
\begin{equation}
\nit{Resp}(\e,F^\star) \ \ := \ \max \limits_{{\Gamma, \bar{w}: \ |\Gamma| \ \\ \mbox{\scriptsize \ is min. \& } \mbox{\scriptsize (\ref{eq:star}) $> 0$}}}
\ \nit{ Resp}(\e,F^\star,\Gamma,\bar{w}). \label{eq:min}
\end{equation}

\vspace{-2mm}
This means that we first find the minimum-size contingency sets $\Gamma$'s for which, for an associated  set of value updates $\bar{w}$, (\ref{eq:star}) becomes greater that $0$. After that, we find the maximum value for (\ref{eq:star}) over all such pairs $(\Gamma,\bar{w})$. This can be done by starting with $\Gamma = \emptyset$, and iteratively increasing the cardinality of $\Gamma$ by one, until a $(\Gamma,\bar{w})$ is found that makes (\ref{eq:star}) non-zero. We stop increasing the cardinality, and we just check if there is another $(\Gamma^\prime, \bar{w}^\prime)$ that gives a greater value for (\ref{eq:star}), with $|\Gamma^\prime| = |\Gamma|$. \
By taking the maximum of the local scores, we  have an existential quantification in mind: there must be a good contingency $(\Gamma, \bar{w})$, as long as $\Gamma$ has a minimum cardinality.
\end{enumerate}

With the generalized score, the difference between counterfactual and actual causes is not as relevant as before. In the end, and as discussed under Item 4. above, what matters is the size of the score. Accordingly, we can talk only about ``counterfactual explanations with responsibility score $r$". In Example \ref{ex:dos}, we could say ``$\e_2$ is a (minimal) counterfactual for $\e$ (implicitly saying that it switches the label), and the value $\msfs{60K}$ for $\msfs{Salary}$ is a counterfactual explanation with responsibility $\nit{Resp}(\e,\msfs{Salary})$". Here, $\e_2$ is possibly only one of those counterfactual entities that contribute to making the value for $\msfs{Salary}$ a counterfactual explanation, and to its (generalized) \Resp \ score.

The generalized $\nit{Resp}$ score was applied for different financial data \cite{deem}, and experimentally compared with  the $\nit{Shap}$ score \cite{LL17,LetA20}, which can also be applied with a black-box classifier, using only the input/output relation. Both were also experimentally compared, with the same data, with a the FICO-score \cite{rudin} that is defined for and applied to an open-box model, and  computes scores by taking into account components of the model, in this case coefficients of nested logistic regressions.


The computation cost of the \Resp \ score is bound to be high in general since, in essence,  it explicitly involves in (\ref{eq:star}) all possible subsets of the set of features; and in (\ref{eq:min}), also the minimality condition which compares different subsets. Actually, for binary classifiers and in its simple, binary formulation, \Resp \ is already intractable \cite{tplp}. \ In \cite{deem}, in addition to experimental results, there is a technical discussion on the importance of the underlying distribution on the population, and on the need to perform optimized computations and approximations.

\vspace{-2mm}
\subsection{The Shap Score and its Tractable Computation}\label{sec:shapTract}

\vspace{-2mm}
The \Shap \ score was introduced in explainable ML in \cite{LL17}, as an application of the general {\em Shapley value} of {\em coalition game theory} \cite{S53}, which we briefly describe next.

Consider a set of players $\mc{S}$,  and a
{\em wealth-distribution  function} (or {\em game function}), $\mc{G}\!: \mc{P}(\mc{S})  \rightarrow  \mathbb{R}$, that maps subsets of $\mc{S}$ to real numbers.  \ The Shapley value of player $p \in \mc{S}$ quantifies the contribution of  $p$ to the game, for which all different coalitions are considered; each time, with $p$ and without $p$:
\vspace{-4mm}

  \begin{equation}\nit{Shapley}(\mc{S},\mc{G},p):= \sum_{S\subseteq
  \mc{S} \setminus \{p\}} \frac{|S|! (|\mc{S}|-|S|-1)!}{|\mc{S}|!}
(\mc{G}(S\cup \{p\})-\mc{G}(S)). \label{eq:shapley}
\end{equation}

\vspace{-2mm} Here, $|S|! (|D|-|S|-1)!$ is the number of permutations of
$\mc{S}$ with all players  in $S$ coming first, then $p$, and then all the others. \ In other words,
this is the {\em expected contribution} of $p$ under all possible additions of $p$ to a partial random sequence of players, followed by random sequences of the rest of the players.

The Shapley value emerges as the only quantitative measure that has some specified properties in relation to coalition games \cite{R88}. It has been applied in many disciplines. \ For each particular application, one has to define a particular and appropriate game function $\mc{G}$. In particular, it has been applied to assign scores to logical formulas to quantify their contribution to the inconsistency of a knowledge base \cite{hunter}, to quantify the contribution of database tuples to making a query true \cite{LBKS20,SigRec21}, and to quantify contributions to the inconsistency of a database \cite{ester}.

In different application and with different game functions, the Shapley  value turns out to be computationally intractable,  more precisely, its time complexity is {\em $\#P$-hard} in the size of the input, c.f., for example, \cite{SigRec21}. Intuitively, this means that it is at least as hard as any of the problems in the class $\#P$ of problems about counting the solutions to decisions problems (in $\nit{NP}$) that ask about the existence of a certain solution \cite{valiant,papadimitriou}. For example, $\nit{SAT}$ is the decision problem asking, for a propositional formula, if there exists a truth assignment (a solution) that makes the formula true. Then, $\#\!\nit{SAT}$ is the computational problem of counting the number of satisfying assignments of a  propositional formula. Clearly, $\#\!\nit{SAT}$ is at least as hard as $\nit{SAT}$ (it is good enough to count the number of solutions to know if the formula is satisfiable), and $\nit{SAT}$ is the prototypical $\nit{NP}$-complete problem, and furthermore, $\#\!\nit{SAT}$ is $\#P$-hard, actually, $\#\!P$-complete since it belongs to $\#P$. As a consequence, computing the Shapley value can be at least as hard as computing the number of solutions for $\nit{SAT}$; a clear indication of its high computational complexity.

As already mentioned, the \Shap \ score is a particular case of the Shapley value in (\ref{eq:shapley}). In this case, the players are the features $F$ in $\mc{F}$, or, more precisely, the values $F(\e)$ they take for a particular entity $\e$, for which we have a binary classification label, $L(\e)$, we want to explain. The explanation takes the form of a numerical score for $F(\e)$, reflecting its relevance for the observed label. Since all the feature values contribute to the resulting label, features values can be seen as players in a coalition game.

 The game function, for a subset $S$ of features, is the {\em expected (value of the) label} over all possible entities whose values coincide with those of $\e$ for the features in $S$:

 \vspace{-2mm}
\begin{equation}
\mc{G}_\mathbf{e}(S) := \mathbb{E}(L(\mathbf{e'})~|~\e^\prime \in \mc{E} \ \mbox{ and } \ \e^{\prime}_{\!S} = \e_S),
\end{equation}

\vspace{-2mm}\noindent where  $\e^\prime_S, \e_S$ denote the projections of $\e^\prime$ and $\e$ on $S$, resulting in two subrecords of feature values. \ We can see that the game function depends on the entity at hand $\e$.

With the game function in (\ref{eq:shapley}), the \Shap \ score for a feature value $F^\star(\e)$ in $\e$ is:

\begin{eqnarray}\nit{Shap}(\mc{F},\mc{G}_\mathbf{e},F^\star) &:=&
\!\!\!\!\!\sum_{S\subseteq
  \mc{F} \setminus \{F^\star\}} \frac{|S|! (|\mc{F}|-|S|-1)!}{|\mc{F}|!}
[ \mathbb{E}(L(\e^\prime|\e^{\prime}_{S\cup \{F^\star\}} = \e_{S\cup \{F^\star\}}) - \nonumber \\
&&\hspace*{4.25cm}\mathbb{E}(L(\e^\prime)|\e^{\prime}_S = \e_S)]. \label{eq:shap}
\end{eqnarray}

\begin{example} \label{ex:tres} (example \ref{ex:dos} cont.) \
For the same fixed entity $\e \ = \ \langle \msfs{john}, \msfs{18}, \msfs{plumber}, \msfs{70K},$ $ \msfs{harlem}, \msfs{10K}, \msfs{basic}\rangle$  and feature $F^\star = \msfs{Salary}$, one of the terms in (\ref{eq:shap}) is obtained by considering $S = \{\msfs{Location}\} \subseteq \mc{F}$: \vspace{-1mm}
\begin{eqnarray*}
&\frac{|1|! (7-1-1)!}{7!} \times
(\mc{G}_\e(\{\msfs{Location}\}\cup \{\msfs{Salary}\})-\mc{G}_\e(\{\msfs{Location}\}))&\\
&\hspace*{3.5cm}= \ \frac{1}{42} \times (\mc{G}_\e(\{\msfs{Location}, \msfs{Salary}\})-\mc{G}_\e(\{\msfs{Location}\})),&
\end{eqnarray*}

\vspace{-1mm}\noindent with, e.g., \ $\mc{G}_\e(\{\msfs{Location}, \msfs{Salary}\})= \mbb{E}(L(\e^\prime)~|~ \e^\prime \in \mc{E}, \  \msfs{Location}(\e^\prime) = \msfs{harlem}, \mbox{ and }$ $ \msfs{Salary}(\e^\prime) = \msfs{70K})$, \  that is, the expected label over all entities that have the same  values as $\e$ for features $\msfs{Salary}$ and $\msfs{Location}$. \ Then, \
$\mc{G}_\e(\{\msfs{Location}, \msfs{Salary}\})-\mc{G}_\e(\{\msfs{Location}\})$ is the expected  difference in the label between the case where the values for $\msfs{Location}$ and $\msfs{Salary}$ are fixed as for $\e$, and the case where only the value for $\msfs{Location}$ is fixed as in $\e$, measuring a local contribution of $\e$'s value for $\msfs{Salary}$.  After that, all these local differences are averaged over all subsets $S$ of $\mc{F}$, and the permutations in which they participate. \boxtheorem
\end{example}

\vspace{-2mm}
We can see that, so as the \Resp \ score, \Shap \ is a {\em local} explanation score, for a particular entity at hand $\e$.  \ Since the introduction of \Shap \ in this form, some variations have been proposed. So as for \Resp,  \Shap \ depends, via the game function, on an underlying probability distribution on the entity population $\mc{E}$. The distribution may impact not only the \Shap \ scores, but also their computation \cite{deem}.

\begin{figure}
\centerline{\includegraphics[width=5cm]{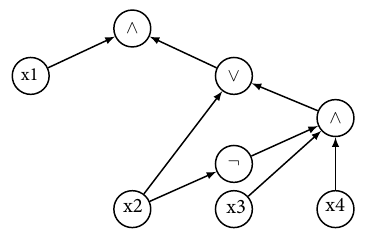}}\vspace{-3mm}
\caption{A decomposable and deterministic Boolean classifier}\label{fig:dDBC}\vspace{-6mm}
\end{figure}

Boolean classifiers, i.e. propositional formulas with binary input features and binary labels, are particularly relevant, {\em per se} and because  they can represent other classifiers by means of appropriate encodings. For example, the circuit in Figure \ref{fig:dDBC} can be seen as a binary classifier that can be represented by means of a propositional formula that, depending on the binary values for $x_1,x_2,x_3,x_4$, also returns a binary value.

Boolean classifiers, as logical formulas, have been extensively investigated. In particular, much is known about the satisfiability problem of propositional formulas, $\nit{SAT}$,  and also about the {\em model counting} problem, i.e. that of counting the number of satisfying assignments, denoted $\#\!\nit{SAT}$.
\ In the area of {\em knowledge compilation}, the complexity of $\#\!\nit{SAT}$ and other problems in relation to the syntactic form of the Boolean formulas have been investigated  \cite{darwicheKC,darwicheJANCL,selman}. \ Boolean classifiers turn out to be quite relevant to understand and investigate the complexity of \Shap \ computation.

The computation of \Shap \ is bound to be expensive, for similar reasons as for \Resp. For the computation of both, all we need is the input/output relation of the classifier, to compute labels for different alternative entities (counterfactuals). However, in principle, far too many combinations have to go through the classifier. Actually, under the {\em product probability distribution} on $\mc{E}$ (which assigns independent probabilities to the feature values), even with an explicit, open classifier for binary entities, the computation of \Shap \ can be intractable.

In fact, as shown in \cite{deem}, for Boolean classifiers in the class $\nit{Monotone}2\nit{CNF}$, of negation-free propositional formulas in conjunctive normal form with at most two atoms per clause, \Shap \ can be $\#\!P$-hard. This is obtained via a polynomial reduction from $\#\nit{Monotone}2\nit{CNF}$, the problem of counting the number of satisfying assignments for a formula in the class, which is known to be $\#\!P$-complete \cite{valiant}. \ For example, if the classifier  is \ $(x_1 \vee x_2) \wedge (x_2 \vee x_3)$, which belongs to $\#\nit{Monotone}2\nit{CNF}$, the entity $\e_1 = \langle 1,0,1\rangle$ (with values for $x_1, x_2, x_3$, in this order) gets label $1$, whereas the entity $\e_2 = \langle 1,0,0\rangle$ gets label $0$. \ The number of satisfying truth assignments, equivalently, the number of entities that get label $1$, is $5$, corresponding to $\langle 1,1,1\rangle$, $\langle 1,0,1\rangle$,  $\langle 0,1,1\rangle$,
 $\langle 0,1,0\rangle$, and $\langle 1,1,0\rangle$.

Given that \Shap \ can be $\#P$-hard, a natural question is whether for some classes of open-box classifiers one can compute \Shap \ in polynomial time in the size of the model and input. The idea is to try to take advantage of the internal stricture and components of the classifier -as opposed to only the input/output relation of the classifier- in order to compute \Shap \ efficiently. We recall from results mentioned earlier in this section that having an open-box model does not guarantee tractability of \Shap. Natural classifiers that have been considered in relation to a possible tractable computation of \Shap \ are decision trees and random forests \cite{LetA20}.

The problem of tractability of \Shap \ was investigated in detail in \cite{AAAI21}, and with other methods in \cite{guyAAAI21ext}. They briefly describe the former approach in the rest of this section.
Tractable and intractable cases were identified, with algorithms for the tractable cases. (Approximations for the intractable cases were further investigated in \cite{AAAI21ext}.) In particular, the tractability for decision trees and random forests was established, which required first identifying the right abstraction that allows for a general proof, leaves aside contingent details, and is also broad enough to include interesting classes of classifiers.

In \cite{AAAI21}, it was proved that, for a Boolean classifier $L$ (identified with its label, the output gate or variable), the uniform distribution on $\mc{E}$, and $\mc{F} = \{F_1, \ldots, F_n\}$: \vspace{-2mm}
\begin{equation}\#\!\nit{SAT}(L) \ = \ 2^{|\mc{F}|} \times (L(\e) - \sum_{i=1}^n \nit{Shap}(\mc{F},G_{\e},F_i)). \label{eq:red}
\end{equation}

\vspace{-3mm}
This result makes, under the usual complexity-theoretic assumptions, impossible for \Shap \ to be tractable for any circuit $L$ for which $\#\!\nit{SAT}$ is intractable. (If we could compute \Shap \ fast, we could also compute $\#\!\nit{SAT}$ fast, assuming we have an efficient classifier.) \ This excludes, as seen earlier in this section, classifiers that are in the class $\nit{Monotone}2\nit{CNF}$. Accordingly, only classifiers in a more amenable class  became candidates, with the restriction that the class should be able to accommodate interesting classifiers. That is how the class of  {\em deterministic and decomposable Boolean circuits} (dDBCs) became the target of investigation.

Each $\vee$-gate of a dDBC can have only one of the disjuncts true (determinism), and for each $\wedge$-gate,
the conjuncts do not share variables (decomposition). Nodes are labeled with $\vee, \wedge$, or $\neg$ gates, and input gates with features (propositional variables) or binary constants. An example of such a classifier, borrowed from \cite{AAAI21},  is shown in Figure   \ref{fig:dDBC}. It has the $\wedge$ gate at the top that returns the output label.

Model counting is known to be tractable for dDBCs. However, this does not imply (via (\ref{eq:red}) or any other obvious way) that \Shap \ is tractable. It is also the case that relaxing any of the determinism or decomposability conditions makes model counting $\#P$-hard  \cite{AAAI21ext}, preventing \Shap \ from being tractable.

It turns out that $\nit{Shap}$ computation is tractable for dDBCs (under the uniform and the product distribution), from which we also get the tractability of $\nit{Shap}$ for free for a vast collection of classifiers that can be efficiently compiled into (or represented as) dDBCs; among them we find:  \ Decision Trees  (even with non-binary features), Random Forests, Ordered Binary Decision Diagrams (OBDDs) \cite{bryant}, Deterministic-Decomposable Negation Normal-Forms  (dDNNFs), Binary Neural Networks  via OBDDs or Sentential Decision Diagrams (SDDs) (but with an extra, exponential, but FPT compilation cost) \cite{darwicheKR20,leon1,leon2}, etc.

\begin{figure}
\centerline{\includegraphics[width=4cm]{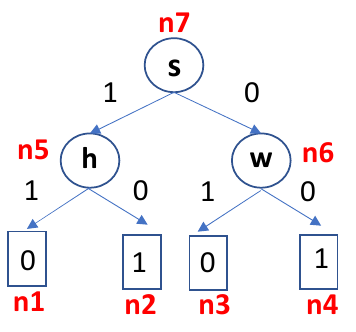}~~~~~~~~~~~~~~~\includegraphics[width=4.5cm]{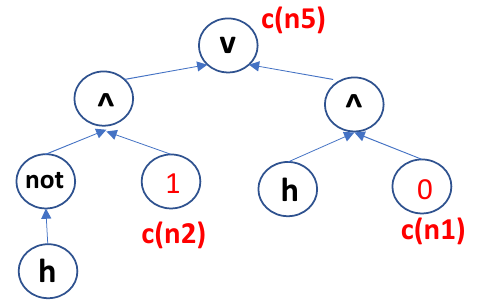}}
\vspace{-2mm}\caption{A decision tree and part of its compilation into an dDBC}\label{fig:dt}
\end{figure}

For the gist, consider the binary decision tree (DT) on the LHS of Figure \ref{fig:dt}. It can be inductively and efficiently compiled into a dDBC \cite[appendix A]{AAAI21ext}. The leaves of the DT
become labeled with propositional constants, $0$ or $1$. Each node, $n$, is compiled into a circuit $c(n)$, and the final dDBC corresponds to the compilation, $c(r)$, of the root node $r$, in this case, $c(n7)$ for node $c7$. \ Figure \ref{fig:dt} shows on the RHS, the compilation $c(n5)$ of node $n5$ of the DT. \ If the decision tree is not binary, it is first binarized, and then compiled \cite[sec. 7]{AAAI21ext}.

\vspace{-7mm}


\vspace{-10mm}

\section{Looking Ahead: Domain Knowledge}\label{sec:last}

\vspace{-2mm}
 There are different approaches and methodologies to provide explanations in data management and artificial intelligence, with causality, counterfactuals and scores being prominent approaches that have a relevant role to play.

Much research is still needed on the use of {\em contextual, semantic and domain knowledge} in explainable data management and explainable machine learning, in particular, when it comes to {\em define and compute} attribution scores. Some approaches may be more appropriate in this direction, and we have argued that declarative, logic-based specifications can be successfully exploited \cite{tplp}. For a particular application, maybe only some counterfactuals make sense, are reasonable or are useful; the latter becoming {\em actionable} or {\em resources} in that they may indicate feasible and achievable conditions (or courses of action) under which we could obtain a different result \cite{ustun,recourse}.

Domain knowledge may come in different forms, among them:
\begin{enumerate}
\item In data management, via semantic constraints, in particular ICs, and ontologies that are fed by data sources or describe them. For example, satisfied ICs might make unnecessary (or prohibited) to consider certain counterfactual interventions on data. In this context, we might want to make sure that sub-instances associated to teams of tuples for Shapley computation satisfy a given set of ICs. Or, at the attribute level, that we do not violate constraints preventing certain attributes from  taking null values.

    \item In the context of machine learning, we could define an {\em entity schema}, basically a wide-predicate $\nit{Ent}(\nit{Feat}_1, \ldots, \nit{Feat}_n)$, on which certain constraints could be imposed. In our running example, we would have the entity schema $\nit{Ent}(\msfs{Name}, \msfs{Age},$ $ \msfs{Activity}, \msfs{Income}, \msfs{Debt}, \msfs{EdLevel})$; and depending on the application domain,  {\em local constraints}, i.e. at the single entity level, such as: \ (a) $\neg(\msfs{Age} < \msfs{6} \wedge  \msfs{EdLevel} = \msfs{phd})$, a denial constraint prohibiting that someone who is less than $6$ years old has a PhD; \ (b) Or an implication \ $\msfs{Activity} = \msfs{none} \rightarrow \msfs{Income} = \msfs{0}$; etc. \ Counterfactual interventions should be restricted by these constraints \cite[sec. 6]{tplp}, or, if satisfied, they could be used to speed up a score computation. In \cite[sec. 7]{tplp}, we showed how the underlying probability distribution, needed for \Resp \ or \Shap,  can be conditioned by logical constraints of this kind.

    We could also have {\em global constraints}, e.g. requiring that any two entities whose values coincide for certain features must have their values for other features coinciding as well. This would be like a functional dependency in databases. This may have an impact of the subteams that are considered for \Shap, for example.

    \item Domain knowledge could also come in the form of {\em probabilistic or statistical constraints}, e.g. about the stochastic independence of certain features, or an explicit stochastic dependency of others via conditional distributions. In this direction, we could have a whole Bayesian network representing (in)dependencies among features. We could also have constraints indicating that certain probabilities are bounded above by a given number; etc. This kind of knowledge would have impact on attribution scores that are defined and computed on the basis of a given probability distribution.
        \end{enumerate}

        The challenge becomes that of bringing these different forms of domain knowledge (and others) into the definitions or the computations of attribution scores.

\ignore{
 Still fundamental research is needed in relation to the notions of {\em explanation} and {\em interpretation}. An always present question is: {\em What is a good explanation?}. \ This is not a new question, and in AI (and other areas and disciplines) it has been investigated. In particular in AI,  areas such as {\em diagnosis} and  {\em causality} have much to contribute.

Now, in relation to {\em explanations scores}, there is still a question to be answered: \ {\em What are the desired properties of an explanation score?}. The question makes a lot of sense, and may not be beyond an answer. After all, the  general
Shapley value emerged from a list of {\em desiderata} in relation to coalition games, as the only measure that satisfies certain explicit properties \cite{S53,R88}. Although the Shapley value is being used in XAI, in particular in its $\nit{Shap}$  incarnation, there could be a different and specific  set of desired properties of explanation scores that could lead to a still undiscovered explanation score.
}

 \vspace{3mm} \noindent {\bf Acknowledgments: }  \ Part of this work was funded by ANID - Millennium Science Initiative Program - Code ICN17002; and NSERC-DG 2023-04650.

\end{document}